# OPTIMIZATION AND NOISE ANALYSIS OF THE QUANTUM ALGORITHM FOR SOLVING ONE-DIMENSIONAL POISSON EQUATION

GUOLONG CUI[1], ZHIMIN WANG[1,*], SHENGBIN WANG[1], SHANGSHANG SHI[1], RUIMIN SHANG[1], WENDONG LI[1], ZHIQIANG WEI[1,2], YONGJIAN GU[1,*]

[1]*College of Information Science and Engineering, Ocean University of China, Qingdao 266100, China*

[2]*High Performance Computing Center, Pilot National Laboratory for Marine Science and Technology (Qingdao), Qingdao 266100, China*

[*]*Correspondence author, e-mail: wangzhimin@ouc.edu.cn; yjgu@ouc.edu.cn*

Solving differential equations is one of the most promising applications of quantum computing. Recently we proposed an efficient quantum algorithm for solving one-dimensional Poisson equation avoiding the need to perform quantum arithmetic or Hamiltonian simulation. In this letter, we further develop this algorithm to make it closer to the real application on the noisy intermediate-scale quantum (NISQ) devices. To this end, we first develop a new way of performing the sine transformation, and based on it the algorithm is optimized by reducing the depth of the circuit from $n^2$ to $n$. Then, we analyze the effect of common noise existing in the real quantum devices on our algorithm using the IBM Qiskit toolkit. We find that the phase damping noise has little effect on our algorithm, while the bit flip noise has the greatest impact. In addition, threshold errors of the quantum gates are obtained to make the fidelity of the circuit output being greater than 90%. The results of noise analysis will provide a good guidance for the subsequent work of error mitigation and error correction for our algorithm. The noise-analysis method developed in this work can be used for other algorithms to be executed on the NISQ devices.

## 1. Introduction

Quantum computer executing quantum algorithms can efficiently solve many problems that are extremely difficult for classical computer [1]. One of the most valuable applications is to solve differential equations, which is the main task in classical high-performance scientific computing [2]. A series of quantum algorithms with exponential speed-up over the classical counterparts have been developed for solving both ordinary and partial differential equations [3-7].

The main idea of most quantum algorithms for solving differential equations is to discretize the differential equations into systems of linear equations, which are then solved by the quantum linear system algorithm (QLSA) [8-9] or Hamiltonian simulation [10-12]. These algorithms aim to solve general differential equations; however, they are too expensive to be implemented using the near-term quantum computers [13-15].

Most recently, we proposed a concise quantum algorithm for solving one-dimensional (1D) Poisson equation (hereafter this algorithm will be called as quantum 1D-Possion solver) [16]. In the spirit of QLSA, our algorithm can encode the solutions of Poisson equation as a quantum state using only $3n$ qubits and $5/3n^3$ elementary gates, where $n$ is the logarithmic of the number of discrete points. This algorithm aims to be implementable on the near-term noisy intermediate-scale quantum (NISQ) devices for real applications [17-18].

The prominent features of NISQ devices are the short coherent time and the noisy quantum gates. The short coherent time limits the depth of the quantum circuit, namely the total number of quantum gates; the operational error of quantum gates will change the amplitude distribution of output states. For a specific circuit representation of a quantum algorithm, it is necessary to study the influence of various noises existing in a real quantum computer on the circuit to evaluate the practicality of actually implementing the algorithm on a quantum computer.

In this paper, we first optimize the quantum 1D-Poisson solver by reducing the depth of the circuit to meet the requirement of limited coherent time. The circuit depth is reduced from $n^2$ to $n$. Secondly, we perform noise analysis for our quantum circuit utilizing both the IBM real quantum computer, IBM SANTIAGO, and the IBM circuit simulator, Qiskit. The main contribution of the present work is the development of a particular noise-analysis method for quantum algorithms to be executed on the NISQ devices.

The paper is organized as follows. First, we provide a high-level overview of the algorithm of quantum 1D-Poisson solver. Next, we optimize the quantum 1D-Poisson solver by redesigning the way of performing the sine transform, and demonstrate the algorithm on the IBMQ_QASM_SIMULATOR. Then, noise analysis of our circuits is performed by combining the IBM real quantum computer and the circuit simulator. Finally, are the conclusions.

## 2. Overview of the quantum 1D-Poisson solver

The one-dimensional Poisson equation with Dirichlet boundary conditions can be expressed as follows,

$$-\frac{d^2v(x)}{dx^2} = b(x), \ x \in (0,1), \quad (1)$$
$$v(0) = v(1) = 0,$$

where $b(x)$ given as input is a smooth function and $v(x)$ is the solution of the equation. Using central difference approximation, Eq. (1) can be discretized into a linear system of equations as follows,

$$h^{-2}(-v_{i-1} + 2v_i - v_{i+1}) = b_i + \varepsilon_i, \ i=1,2,...,N-1 \quad (2)$$
$$v_0 = v_N = 0,$$

where $h=1/N$ is mesh size and the number of discrete points is $N+1$. Ignore the truncation error $\varepsilon_i$, then Eq. (2) turns to be

$$A\vec{v} = \vec{b} \Leftrightarrow h^{-2} \begin{pmatrix} 2 & -1 & & 0 \\ -1 & \ddots & \ddots & \\ & \ddots & \ddots & -1 \\ 0 & & -1 & 2 \end{pmatrix} \begin{pmatrix} v_1 \\ v_2 \\ \vdots \\ v_{N-1} \end{pmatrix} = \begin{pmatrix} b_1 \\ b_2 \\ \vdots \\ b_{N-1} \end{pmatrix}. \quad (3)$$

The coefficient matrix $A$ is a tridiagonal Toeplitz matrix whose eigenvectors and corresponding eigenvalues are $u_j(k) = \sqrt{2/N}\sin(j\pi k/N)$ and $\lambda_j = 4N^2\sin^2(j\pi/2N)$.

The quantum version of solving Eq. (3) is to produce such a quantum state that

$$|v\rangle = A^{-1}|b\rangle = (\sum_j \frac{1}{\lambda_j}|u_j\rangle\langle u_j|)\cdot(\sum_{j'}\beta_{j'}|u_{j'}\rangle) = \sum_j C\frac{\beta_j}{\lambda_j}|u_j\rangle, \qquad (4)$$

where $C$ is the normalizing constant. In Ref. [16] we show that such a state can be prepared in a very simple way, which avoid the need of invoking the high-cost subroutines of QLSA or Hamiltonian simulation. The main point is that the reciprocal of eigenvalues of matrix $A$ can be computed according to the following equation,

$$\frac{8}{\lambda_j} = \left[\left(\sin\frac{\pi}{6}\right)^m \prod_{k=2}^{n-m}\sin(\frac{|2^k - 2^{-m}j \bmod 2^{k+1}|}{2^{k+1}}\pi)\right]^2. \qquad (5)$$

That is, each $1/\lambda_j$ in Eq. (4) can be calculated using $n$-1 terms of square of sine values, where $n$ equals $\log(N)$. Furthermore, the sine-square terms can be prepared easily by a series of single-qubit $R_y$ rotations.

The overall quantum circuit to solve the one-dimensional Poisson equation is shown in Figure 1. The sine transform (ST) is used to change the basis of Register B from eigenstates $|u_j\rangle$ to computational basis $|j\rangle$, namely $\sum_j \beta_j|u_j\rangle \to \sum_j \beta_j|j\rangle$. Then the reciprocals of eigenvalues are calculated under the control of computational basis $|j\rangle$ through a series of $R_y$ rotations according to Eq. (5). Details of this part can be found in Ref. [16]. After the multi-controlled CNOT operation and the inverse sine transform (ST$^\dagger$), the state evolves to

$$\sum_{j=1}^{2^n-1}\beta_j\left[\sqrt{1-\left(\frac{8}{\lambda_j}\right)^2}|\gamma\rangle_E|0\rangle + \frac{8}{\lambda_j}|1\rangle_E^{\otimes 2n-2}|1\rangle\right]|u_j\rangle_B, \qquad (6)$$

where $|\gamma\rangle_E$ represents all the states with at least one qubit being $|0\rangle$ in register E. Finally, the ancilla qubit is measured, and if the result is $|1\rangle$, the solution state of the one-dimensional Poisson equation is created successfully in register B. Before the measurement, amplitude amplification is usually implemented to increase the success probability of obtaining the expected state.

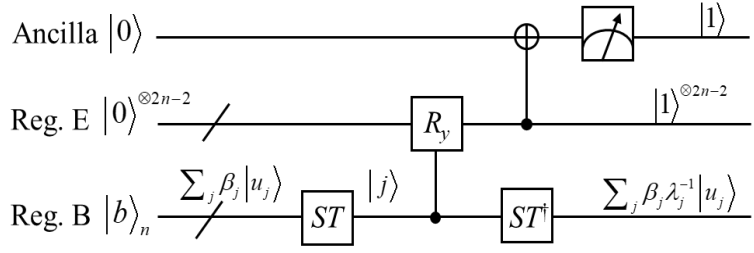

Figure 1 The overall circuit for solving one-dimensional Poisson equation. Register B encodes the input states $|b\rangle = \sum_j \beta_j |u_j\rangle$, and ST represents the sine transform. Register E is used to compute the reciprocals of eigenvalue through a series of $R_y$ rotations. Finally, Ancilla qubit will be measured with the success flag being $|1\rangle$.

The complexity of the circuit in Figure 1 is $3n$ in qubits and $5/3n^3$ in elementary gates, which is rather low comparing with the algorithms in Refs. [19,24]. Here we find that the cost can be reduced further by changing the way of implementing the sine transform and combining it properly with the following series of $R_y$ rotations. In the next section, we will show the optimized quantum 1D-Poisson solver.

## 3. Optimization and demonstration of the quantum 1D-Poisson solver

### 3.1. Optimize the quantum 1D-Poisson solver

We optimize the quantum 1D-Poisson solver by redesigning the way of performing the sine transform (ST). The sine transform corresponds to a matrix $\{ST_{i,j}\}_{i,j=1,...,N-1}$ with $ST_{i,j} = \sqrt{2/N}\sin(\pi ij/N)$. Generally, the sine transform can be implemented through the Fourier transform (FT) as follows,

$$T_{2N}^{\dagger} FT_{2N} T_{2N} = \begin{pmatrix} CT_{N+1} & 0 \\ 0 & -i \cdot ST_{N-1} \end{pmatrix}, \qquad (7)$$

where $CT_{N+1}$ ($ST_{N-1}$) represents the cosine (sine) transform and the subscript denotes the size of the corresponding matrix [20]. As shown in Figure 2, the sine transform can be extracted easily from the matrix of $T_{2N}^{\dagger} FT_{2N} T_{2N}$ by utilizing one ancilla qubit. The quantum Fourier transform can be implemented efficiently in an exact way [21] or an approximated way [22]. Below we discuss the $T_{2N}$ transformation, which is the starting point we improve the way of implementing the sine transform.

The general quantum circuits for implementing the unitary $T_{2N}$ are discussed in Refs. [23]. The $T_{2N}$ transformation mainly consists of an add-one operation, which is usually implemented in a way as shown in Figure 3(a) [24]. The advantage of this way is that it does not need any ancillary qubits, while the disadvantage is the heavy cost of operations and circuit depth mainly resulting from the multi-controlled NOT gates.

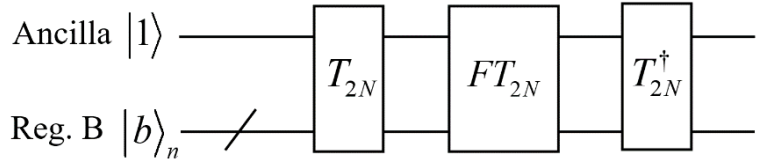

Figure 2 The frame to implement the sine transform. The ancilla qubit initialized to $|1\rangle$ is used to pick out the sine transform from the transformation matrix $T_{2N}^{\dagger} FT_{2N} T_{2N}$.

In the present work, we adapt the quantum ripple carry adder [25] to implement the add-one operation as shown in Figure 3(b). The carry adder can reduce the depth of the add-one circuit greatly, that is, the $n$ $n$-controlled NOT gates are replaced by $n$ TOFFOLI gates as shown in the figure. Specifically, using the new way of performing addition-one operation, the sine transform requires at most $3n^2+2n-1$ basic gates and the depth of the circuit is $2n$. As comparison, the previous method needs at least $4n^2-n+2$ gates, and the depth is at least $n^2+3n+1$ according to Ref. [27]. This improvement brings the quantum 1D-Poisson solver closer to the real application on quantum computer in the near future.

Note that carry adder needs $n-1$ additional ancillary qubits, but it will not increase the total qubit cost of the whole circuit of solving Poisson equation, because these ancillary qubits (i.e. the carry register C in Figure 3(b)) are reversed to zero and they can be reused in the following $R_y$ rotation operations as discussed below.

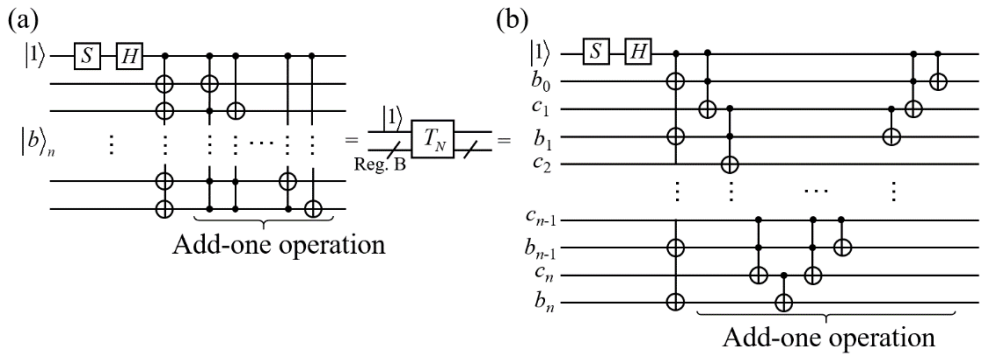

Figure 3 The quantum circuits for implementing $T_{2N}$ transformation before (a) and after (b) optimization. In the optimized circuit (b), the ripple carry adder is used to perform add one operation, and an additional register C (i.e. $c_1$, $c_2...c_n$) is used to stores the carry.

Combine the new sine transform with the following $R_y$ rotations, and then we obtain the optimized quantum 1D-Poisson solver as shown in Figure 4. Comparing with the previous algorithm as shown in Figure 1, the main difference is that the Register E in Figure 1 is divided into two registers in Figure 4. One of the registers, namely Register C, is first used to implement the sine transformation before reversing the eignvalues.

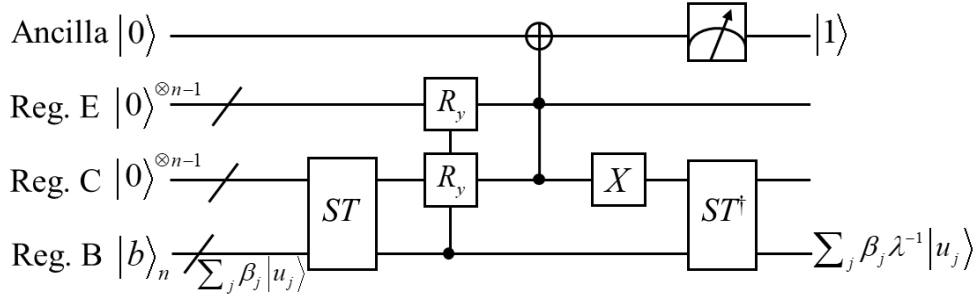

Figure 4 The optimized quantum circuit for solving one-dimensional Poisson equation. Register B and Ancilla are the same as those in Figure 1. Register C is the carry register in Figure 3. Register C and E is combined to implement the series of $R_y$ rotations.

Now let us sketch how the quantum states evolve through the circuit in Figure 4. As before, first the sine transform change the basis in Register B. After the implement of the $R_y$ rotations and controlled NOT operations, the state evolves into

$$\sum_{j=1}^{2^n-1} \beta_j \left[ \sqrt{1-\left(\frac{8}{\lambda_j}\right)^2} |\gamma\rangle_{E\&C} |0\rangle + \frac{8}{\lambda_j} |1\rangle_E^{\otimes n-1} |1\rangle_C^{\otimes n-1} |1\rangle \right] |j\rangle, \qquad (8)$$

where $|\gamma\rangle_{E\&C}$ represents all the states with at least one qubit being $|0\rangle$ in register E and C. Next, the $X$ module flip the states of Register C qubits, i.e. the second term in equation (8) from $|1\rangle$ to $|0\rangle$ as follows,

$$\sum_{j=1}^{2^n-1} \beta_j \left[ \sqrt{1-\left(\frac{8}{\lambda_j}\right)^2} |\gamma\rangle_{E\&C} |0\rangle + \frac{8}{\lambda_j} |1\rangle_E^{\otimes n-1} |0\rangle_C^{\otimes n-1} |1\rangle \right] |j\rangle. \qquad (9)$$

Next step is to inverse the sine transform, which transforms the state into

$$\sum_{j=1}^{2^n-1} \beta_j \sqrt{1-\left(\frac{8}{\lambda_j}\right)^2} ST^\dagger |j\rangle |\gamma\rangle_{regE\&C} |0\rangle + \sum_{j=1}^{2^n-1} \beta_j \frac{8}{\lambda_j} |u_j\rangle |1\rangle_{regE}^{\otimes n-1} |0\rangle_{regC}^{\otimes n-1} |1\rangle. \qquad (10)$$

Before measurement, we can perform amplitude amplification to increase the success probability of obtaining the targeted state. Finally, we measure the ancilla qubit; if the measurement result is $|1\rangle$, then the solution state to the one-dimensional Poisson equation is created successfully in Register B.

*3.2. Demonstration of the optimized quantum 1D-Poisson solver*

We take the cases of *n*=2 and *n*=3 for example to demonstrate the optimized quantum 1D-Poisson solver. In the cases of *n*=2, the discretized matrix of the Poisson equation is 3×3 corresponding to 5 discretized points, while for *n*=3 the matrix is 7×7 with 9 points. Figure 5 shows the circuit for *n*=2.

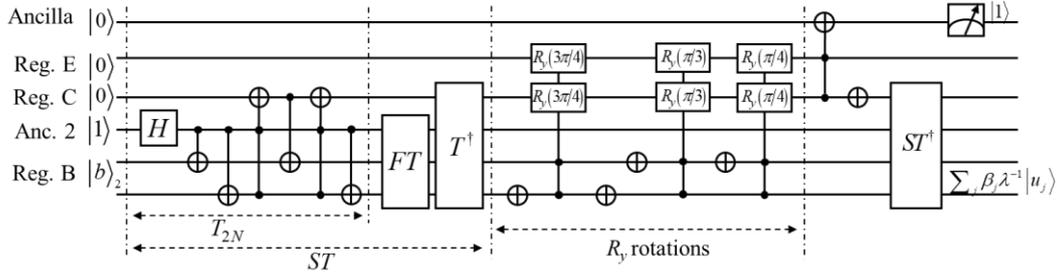

Figure 5 The demonstrated circuit for quantum 1D-Poisson equation with *n*=2. The Anc. 2 qubit initialized in $|1\rangle$ corresponds to the ancillary qubit in Figure 3. The NOT gate before the ST$^\dagger$ transform is the X module in Figure 4. The circuit for n=3 can be obtained accordingly.

We use the IBM quantum-circuit simulator, QASM_SIMULATORv0.1.547 [17], to execute the circuits of *n*=2 and *n*=3. The initial state $|b\rangle_2$ for *n*=2 and $|b\rangle_3$ for *n*=3 are prepared as $1/\sqrt{2}|01\rangle+1/2|10\rangle+1/2|11\rangle$ and $1/2|001\rangle+\sqrt{2}/4|010\rangle+\sqrt{2}/4|011\rangle+\sqrt{2}/4|100\rangle+\sqrt{2}/4|101\rangle$ $\sqrt{2}/4|110\rangle+\sqrt{2}/4|111\rangle$ respectively. In theory, probabilities of the computational basis {01, 10, 11} are {0.205, 0.304, 0.161} for *n*=2 circuit and {0.029, 0.078, 0.118, 0.132, 0.114, 0.073, 0.025} for basis {001, 010, 011, 100, 101, 110, 111} of *n*=3 circuit. Note that the above probabilities of the computational basis are the ones when the state of Ancilla in Figure 5 is measured as $|1\rangle$, so the summation of them would not be 1.

The simulation results for n=2 and n=3 circuits are shown in Figure 6. The codes for the two circuits can be found in the Appendix A. The codes are executed on the IBM Qiskit open-source software. They are written in an intelligible quantum assembly language developed by the IBM. The simulation results are in good coincidence with the theoretical ones, and thus the optimized quantum 1D-Poisson solver is verified.

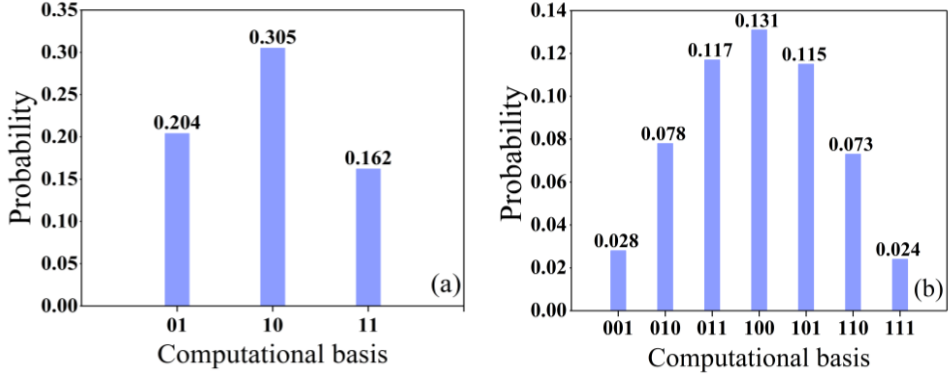

Figure 6 The simulation results of the optimized quantum 1D-Poisson solver for $n=2$ (a) and $n=3$ (b) using the IBM circuit simulator, Qiskit. The computational basis is that of Reg. B in Figure 5. The results coincidence well with the theoretical ones.

## 4. Noise analysis of the optimized quantum 1D-Poisson solver

*4.1. Noise in the real quantum computer*

Noise existing in the quantum computer is from the interaction between quantum systems and environments. Every completely positive trace-preserving map $\varepsilon$ can be regarded as a channel which can be represented in the Kraus form [26],

$$\rho' = \varepsilon(\rho) = \sum_k E_k \rho E_k^\dagger, \quad (11)$$

where $\rho$ is the density matrix of initial state $\rho'$ is density matrix of final state after dynamic process and $E_k$ represents a series of Kraus operators. It describes the dynamic evolution of the quantum system. Different noise can be described by a particular series of Kraus operators.

In general, there are four common kinds of noise, that is, amplitude damping, phase damping, bit flip and depolarizing noise. The amplitude damping noise can be used to describe the loss of energy from the quantum system to the environment, such as the process of photon emission. The Kraus operators $E_k$ corresponding to amplitude damping are [21]

$$E_0 = \begin{bmatrix} 1 & 0 \\ 0 & \sqrt{1-p} \end{bmatrix}, \quad E_1 = \begin{bmatrix} 0 & \sqrt{p} \\ 0 & 0 \end{bmatrix}, \quad (13)$$

where $p (0 \le p \le 1)$ is the probability of a quantum state occurring error. The Kraus operator $E_0$ leaves the state $|0\rangle$ unchanged and reduces the amplitude of state $|1\rangle$, while $E_1$ changes the state $|1\rangle$ into $|0\rangle$ corresponding to the process of losing energy for the quantum system. Note that a quantum state occurring error means that there appears noise in the circuit, so the noise intensity is represented by the magnitude of the probability $p$. This relationship will be used below.

The phase damping noise describes the loss of quantum information of a quantum system, but without loss of energy, like the process of photon scattering [21]. Its Kraus operators are

$$E_0 = \begin{bmatrix} 1 & 0 \\ 0 & \sqrt{1-p} \end{bmatrix}, \quad E_1 = \begin{bmatrix} 0 & 0 \\ 0 & \sqrt{p} \end{bmatrix}, \tag{14}$$

where $p(0 \leq p \leq 1)$ is the probability of a quantum state occurring error, i.e. the probability of occurring photons scattering (without loss of energy). The Kraus operator $E_0$ leaves the state $|0\rangle$ unchanged and reduces the amplitude of the state $|1\rangle$, while $E_1$ destroys state $|0\rangle$ and reduces the amplitude of state $|1\rangle$.

The bit flip noise changes the state of a qubit from $|0\rangle$ to $|1\rangle$ and vice versa with a probability of $1-p$, and leave the state unchanged with probability $p$. So the Kraus operators is as follows [21],

$$E_0 = \sqrt{p}\begin{bmatrix} 1 & 0 \\ 0 & 1 \end{bmatrix}, \quad E_1 = \sqrt{1-p}\begin{bmatrix} 0 & 1 \\ 1 & 0 \end{bmatrix}. \tag{15}$$

The depolarizing channel depolarizes a state of qubit into a completely mixed state $I/2$ with probability $p$, and with a probability $1-p$ the state is left unchanged. The Kraus operators is as follows [21],

$$E_0 = \sqrt{1-\frac{3p}{4}}\begin{bmatrix} 1 & 0 \\ 0 & 1 \end{bmatrix}, E_1 = \frac{\sqrt{p}}{2}\begin{bmatrix} 0 & 1 \\ 1 & 0 \end{bmatrix}, E_2 = \frac{\sqrt{p}}{2}\begin{bmatrix} 0 & -i \\ i & 0 \end{bmatrix}, E_3 = \frac{\sqrt{p}}{2}\begin{bmatrix} 1 & 0 \\ 0 & -1 \end{bmatrix}. \tag{16}$$

In order to analyze the noise effect on the quantum circuit, one needs to place the Kraus operators before each quantum gate with certain probability, and analyze how the quantum states are changed along with the probability.

*4.2. Noise analysis of the quantum 1D-Poisson solver*

We perform the noise analysis through three steps: first benchmarking the noise module provided by the IBM Qiskit, then analyzing the effect of the four kinds of noise discussed above respectively, finally quantifying the noise effect on the quantum 1D-Poisson solver circuit.

To start, the quantum circuits are executed in three ways, i.e. by the ideal circuit simulator without noise, by the circuit simulator with device backend noise model provided in the IBM Qiskit, and by the real quantum computer IBMQ_SANTIAGO (five qubits) [17]. The obtained results in the three scenarios for *n*=2 circuit are compared in Figure 7. The running codes are given in the Appendix A.

As can be seen from the Figure 7, the noise existing in the real quantum computer indeed has a great effect on the output states. In fact, the noises nearly smooth the characteristic amplitude distribution of the computational basis of the output states. The deviation would be up to 80 percent. That is, the circuit fail to encode the solutions of the 1D-Poisson equation into its output states. In addition, we can see that the device backend noise model in the IBM Qiskit can simulate, to some extent, the effect of the noise in the real quantum device. This noise model is a good approximation of the real noise, which can provide us a good reference and starting point to analyze the noise effect in more detail. In fact, the device backend noise model is generated using the calibration information of IBM's real device, which can mimic a real quantum computer approximately (here we chose IBMQ_SANTIAGO).

The backend noise model includes a single qubit depolarizing error followed by a single qubit thermal relaxation error, a two-qubit depolarizing error followed by single-qubit thermal relaxation errors, and the single-qubit readout errors when measuring individual qubit.

However, the device backend noise model provided by the IBM Qiskit can only be used to simulate the total effect of all kinds of noise. Moreover, the maximum number of bits it supports is equal to that of real device it mimics. In order to get more detail about the noise effect on our quantum 1D-Poisson solver circuit, we individually add the four common noises in section 4.1 into the circuit with reasonable error value (i.e. the noise intensity). When adding a noise to a quantum gate, the corresponding Kraus operators of the noise is placed before the logic gate.

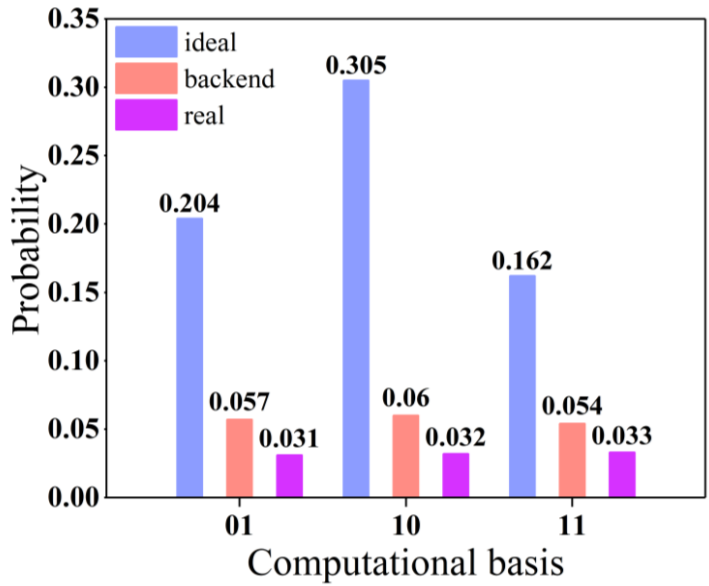

Figure 7 The execution results of the quantum 1D-Poisson solver circuit (with $n=2$) by ideal circuit simulator without noise, circuit simulator with device backend noise model, and real quantum computer. The real quantum computer is the IBMQ_SANTIAGO. All the probability values in the figure are the average over three sets of data and the same is true for the following figures.

We add the four kinds of noise into the $n=2$ and $n=3$ circuits of the quantum 1D-Poisson solver. The IBM Qiskit toolkit is employed to accomplish this task. It contains classes and functions to help building desired noise model to simulate a quantum circuit. The simulation results for the $n=2$ circuit is shown in Figure 8. The corresponding code is given in Appendix A. The noise intensity, namely the magnitude of the probability $p$ in Kraus operators, is set to be $1.8\times10^{-2}$ according to the real parameters of the 5-qubit IBMQ_SANTIAGO quantum computer.

It is interesting to see from Figure 8 that the noise of phase damping has small effect on the circuit output. This is due to the fact that our algorithm is designed to evolve the quantum states on the amplitudes that the phase damping would has little effect on the output states. The other three kinds of noise, the amplitude damping, bit flip and depolarizing channel, has similar effect on the circuit. This conclusion is useful, which tell us that when performing error correction, the phase damping error can be ignored with small loss of accuracy. Furthermore, it shows that when considering the four kinds of noise, we can properly mimic the effect of the noise existing in the real quantum device.

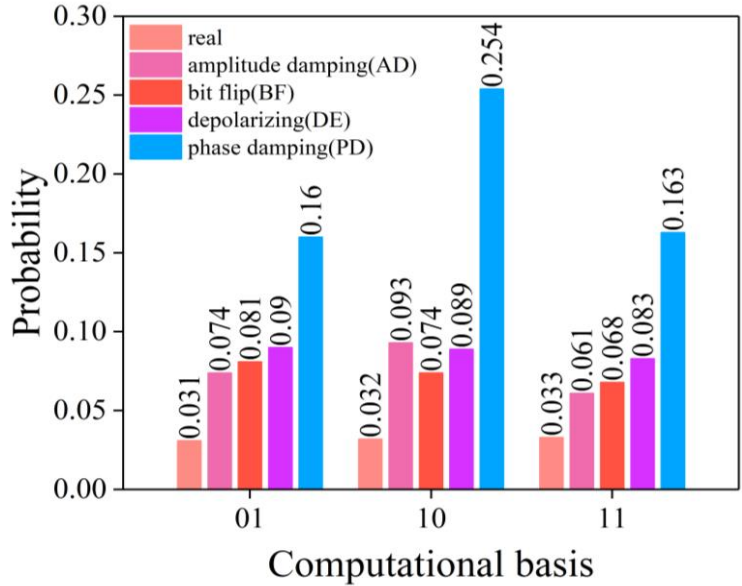

Figure 8 The running results of the $n=2$ quantum 1D-Poisson solver on the IBM real quantum computer, as well as the numerical simulation results under the effect of four types of noise. For each computational basis, the order of the data from left to right is the same as that in the legend from up to down.

In the last step, we quantify the noise effect on our quantum 1D-Poisson solver. The four kinds of noise are added into $n=2$ and $n=3$ circuits with different order of magnitudes of noise intensity. Specifically, the probability $p$ in Kraus operators for each kind of noise is set to be changed from $10^{-4}$ to $10^{-2}$ step-by-step according to the following equation [28],

$$p_i = 10^{-4} \times 700^{0.1 \times i}, \quad i = 1, 2, ..., 9. \tag{17}$$

In addition, the deviation of the probability of each computational basis of the output states is taken as a measure to quantify the noise effect. It is defined as

$$\bar{D} = E_{basis}(D), \quad D = \left| \frac{p^{noise} - p^{theory}}{p^{theory}} \right|, \tag{18}$$

where $p^{noise}$ denotes the measured probability of the target computational basis with a given noise type and level, and $p^{theory}$ is the theoretical one without noise effect. The $E_{basis}(D)$ represents the average of the deviation $D$ over all targeted computational basis.

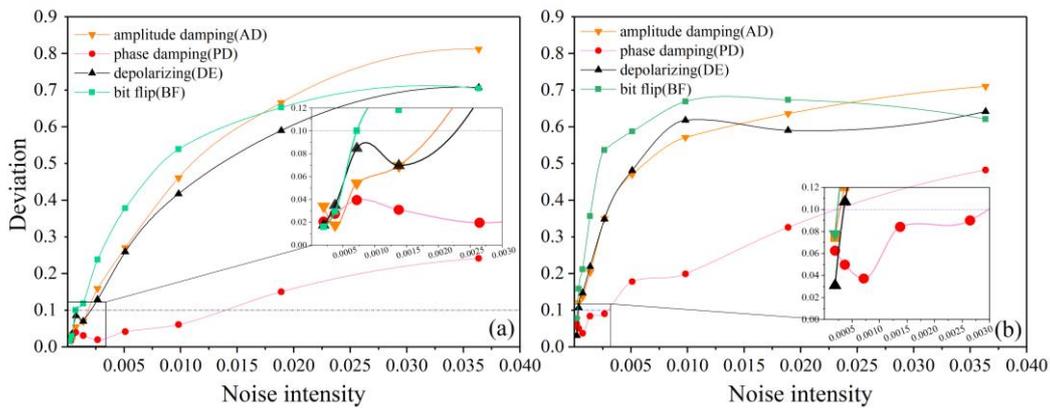

Figure 9 The deviation of the probability of targeted computational basis along with the increasing of noise intensity for the $n=2$ (a), and $n=3$ (b) 1D-Poisson solver circuit. The noise intensity is set according to Eq. (17). The specific data can be found in Table 1 and 2 in the Appendix B.

The simulation results of the deviation $\bar{D}$ for $n=2$ and $n=3$ circuit are shown in Figure 9. (The running codes are given in the Appendix A.) Obviously, as can be seen from the figure, along with the error decreasing, the deviation become smaller and smaller. In order to guarantee that the deviation of the amplitude distribution is below 10%, the error of the quantum gates should be lower than $7.1 \times 10^{-4}$ for the $n=2$ circuit with 131 basic quantum gates; while for the $n=3$ circuit with 361 basic gates, the error should lower than $1.9 \times 10^{-4}$. This result also provides a good quantitative reference for the noise effect on NISQ algorithms whose number of gates are similar or proportional to the above two circuits. In addition, we find that when the noise level is moderate (neither too imperceptible nor too noisy, which is also the typical noise level of the present NISQ devices), the bit flip noise will cause the largest deviation for the probability distribution of the output states. Furthermore, the four types of noise can be added to the quantum circuit simultaneously to approximate the noise situation in the real quantum computers.

## 5. Conclusions

In the present work, we optimize the quantum algorithm for solving the one-dimensional Poisson equation by reducing the circuit depth from $n^2$ to $n$. To achieve the optimization, we develop a new way of implementing the sine transformation, which would be useful for other quantum algorithms. The optimized quantum 1D-Poisson solver is demonstrated using the IBM circuit simulator.

In order to make the quantum 1D-Poisson solver closer to the real applications on the NISQ devices, we analyze the noise effect of four common kinds of noise on the circuits. We first execute the circuits successfully on a 5-qubits real quantum computer, the IBM_ SANTIAGO, and show the great effect of the noise on the circuit output states. Then we quantify the noise effect of the four kinds of noise, including the amplitude damping, phase damping, bit flip and depolarizing noise, using the IBM Qiskit toolkit. We find that the phase damping noise has little effect on our algorithm, while the bit flip noise causes the greatest effect in the concerned cases. In addition, for $n=2$ and $n=3$ quantum 1D-Poisson solver circuits with 131 and 361 basic gates, the threshold error of the quantum gates should be lower than $7.1 \times 10^{-4}$ and $1.9 \times 10^{-4}$, respectively, to make the deviation of the output probability distribution below 10%. These results provide a good guidance for our subsequent work to implement error mitigation and error correction on our quantum 1D-Poisson solver.

**Acknowledgements**

We acknowledge the use of IBM Quantum services for this work. The views expressed are those of the authors, and do not reflect the official policy or position of IBM or the IBM Quantum team. The present work is supported by the National Natural Science Foundation of China (Grant No. 12005212, 6157518) and the Pilot National Laboratory for Marine Science and Technology (Qingdao).

**References**
1. Shor P W (1994), *Algorithms for quantum computation: discrete logarithms and factoring*, Proceedings 35th annual symposium on foundations of computer science, pp. 124-134
2. Harrow A W, Hassidim A and Lloyd S (2009), *Quantum algorithm for linear systems of equations*, Phys. Rev. Lett. 103, pp. 150502
3. Leyton S K and Osborne T J (2008), *A quantum algorithm to solve nonlinear differential equations*, arXiv:0812.4423
4. Clader B D, Jacobs B C and Sprouse C R (2013), *Preconditioned quantum linear system algorithm* ,Phys. Rev. Lett. 110, pp. 250504
5. Berry D W, Childs A M, Ostrander A and Wang, G (2017), *Quantum algorithm for linear differential equations with exponentially improved dependence on precision*, Commun.

Math. Phys. 356, pp. 1057-1081

6. Childs A M and Liu J P (2020), *Quantum spectral methods for differential equations*, Commun. Math. Phys. 375, pp. 1427-1457
7. Arrazola J M, Kalajdzievski T, Weedbrook C and Lloyd S (2019), *Quantum algorithm for nonhomogeneous linear partial differential equations*, Phys. Rev. A 100, pp. 032306.
8. Berry D W (2014), *High-order quantum algorithm for solving linear differential equations*, J. Phys. A-Math. Theor. 47, pp. 105301
9. Childs A M, Kothari R and Somma R D (2017), *Quantum algorithm for systems of linear equations with exponentially improved dependence on precision*, SIAM J. Comput. 46, pp. 1920-1950
10. Berry D W, Childs A M and Kothari R (2015), *Hamiltonian simulation with nearly optimal dependence on all parameters*, IEEE 56th Annual Symposium on Foundations of Computer Science, pp. 792-809
11. Berry D W, Childs A M, Cleve R, Kothari R and Somma R D (2014), *Exponential improvement in precision for simulating sparse Hamiltonians*, Proceedings of the forty-sixth annual ACM symposium on Theory of computing, pp. 283-292
12. Low G H and Chuang I L (2017), *Optimal Hamiltonian simulation by quantum signal processing*, Phys. Rev. Lett. 118, pp. 010501
13. Arute F, Arya K, Babbush R, Bacon D., Bardin J C, Barends R, ... and Martinis J M (2019), *Quantum supremacy using a programmable superconducting processor*, Nature 574, pp. 505-510
14. Jurcevic P, Javadi-Abhari A, Bishop L S, Lauer, I, Bogorin D F, Brink M, ... and Gambetta J M (2021), *Demonstration of quantum volume 64 on a superconducting quantum computing system*, Quantum Sci. Technol. 6, pp. 025020
15. Sundaresan N, Lauer I, Pritchett E, Magesan E, Jurcevic P, and Gambetta J M (2020), *Reducing unitary and spectator errors in cross resonance with optimized rotary echoes*, PRX Quantum 1, pp. 020318
16. Wang S, Wang Z, Li W, Fan L, Cui G, Wei Z, and Gu, Y (2020), *A quantum Poisson solver implementable on NISQ devices*, arXiv:2005.00256
17. See IBM Quantum. https://quantum-computing.ibm.com/ (2021) for detailed information
18. Gong M, Wang S, Zha C, Chen M C., Huang H L, Wu Y, ... and Pan J W (2021), *Quantum walks on a programmable two-dimensional 62-qubit superconducting processor*, Science 372, pp. 948-952
19. Wang S, Wang Z, Li W, Fan L, Wei Z, and Gu Y (2020), *Quantum fast Poisson solver: the algorithm and complete and modular circuit design*, Quantum Inf. Process. 19, pp. 1-25
20. Wickerhauser M V. (1996), *Adapted wavelet analysis: from theory to software* (CRC: Boca Raton), pp. 68-90
21. Nielsen M A and Chuang I. (2002), *Quantum computation and quantum information* (Cambridge University: New York)
22. Yunseong N, Yuan Su, and Dmitri M (2020), *Approximate quantum Fourier transform with O (n log(n)) T gates*, npj Quantum Inf. 6, pp. 26
23. Klappenecker A and Rotteler M (2001), *Discrete cosine transforms on quantum computers*, Proceedings of the 2nd International Symposium on Image and Signal Processing and Analysis. In conjunction with 23rd International Conference on Information Technology Interfaces, pp. 464-468
24. Cao Y, Papageorgiou A, Petras I, Traub J, and Kais S (2013), Quantum algorithm and circuit design solving the Poisson equation, New J. Phys. 15, pp. 013021
25. Vedral V, Barenco A and Ekert A. (1996), *Quantum networks for elementary arithmetic operations*, Phys. Rev. A 54, pp. 147
26. Xian-Ting L (2003), Classical information capacities of some single qubit quantum noisy channels, Commun. Theor. Phys. 39, pp. 537
27. Barenco A, Bennett C H, Cleve R, DiVincenzo D P, Margolus N, Shor P, ... and Weinfurter H (1995), *Elementary gates for quantum computation*, Phys. Rev. A 52, pp. 3457

28. Xue C, Chen Z Y, Wu Y C and Guo G P (2021), *Effects of quantum noise on quantum approximate optimization algorithm*, Chin. Phys. Lett. 38, pp. 030302

**Appendix A CODE**

Here we provide the Qiskit codes for the case of *n*=2 and 3. In the section first in order to visualize the designed circuit the code is taken the form in IBM online computing platform which can be transformed into Qiskit code easily by drop-down window on the interface [1]. Then The Qiskit code for ideal circuit simulation without noise, for device backend noise model mimicked IBMQ_SANTIAGO, for the real quantum computer IBMQ_SANTIAGO, for adding four common kinds of noise are given in turn. The execution of Qiskit code depends on the Python environment. So, first, we need install the software Anaconda, next in Anaconda prompt create a python environment. Then basing on the environment build a Qiskit package [2]. Finally typing "python" to start the simulation.

**The circuit code for n=2**

```
#the simulator has its basic gates of H, S, X, T, CNOT, TOFFOLI, U, U2, U3, etc
#they are converted to one & two qubit gates to get the number of 79
#5 qubit omitted the auxiliary bits in fig.5 instead of observing Reg. E and Reg. C directly
#qubit 0 and 2 corresponds to Reg. E and      C respectively
#qubit 3 and 1 Reg. B
#qubit 4 Anc.2

qreg q[5];
creg c[4];

#state b preparation
h q[3];
x q[4];
x q[3];
h q[4];
cx q[3],q[1];
x q[3];
ch q[3],q[1];
#state b preparation done

#sin transform
cx q[4],q[3];
cx q[4],q[1];
ccx q[4],q[1],q[2];
cx q[2],q[3];
ccx q[4],q[1],q[2];
cx q[4],q[1];
h q[4];
cu1(pi/2) q[3],q[4];
cu1(pi/4) q[1],q[4];
h q[3];
cu1(pi/2) q[1],q[3];
h q[1];
swap q[1],q[4];
cx q[4],q[1];
ccx q[4],q[1],q[2];
cx q[2],q[3];
ccx q[4],q[1],q[2];
cx q[4],q[1];
cx q[4],q[3];
#sin transform done

#controlled R_y module
x q[3];
ccx q[3],q[1],q[2];
ccx q[1],q[3],q[0];
cry(pi/8) q[1],q[2];
cry(pi/8) q[1],q[0];
cx q[3],q[1];
cry(-pi/8) q[1],q[2];
cry(-pi/8) q[1],q[0];
cx q[3],q[1];
cry(pi/8) q[3],q[2];
cry(pi/8) q[3],q[0];
x q[1];
x q[3];
cry(pi/6) q[1],q[2];
cry(pi/6) q[1],q[0];
cx q[3],q[1];
cry(-pi/6) q[1],q[2];
cry(-pi/6) q[1],q[0];
cx q[3],q[1];
cry(pi/6) q[3],q[2];
cry(pi/6) q[3],q[0];
x q[1];
cry(pi/8) q[1],q[2];
cry(pi/8) q[1],q[0];
```

```
cx q[3],q[1];
cry(-pi/8) q[1],q[2];
cry(-pi/8) q[1],q[0];
cx q[3],q[1];
cry(pi/8) q[3],q[2];
cry(pi/8) q[3],q[0];
#controlled Ry module done

x q[2]; #X module in Fig. 4
cx q[4],q[3];
cx q[4],q[1];
ccx q[4],q[1],q[2];
cx q[2],q[3];
ccx q[4],q[1],q[2];
cx q[4],q[1];
swap q[1],q[4];
h q[1];
cu1(-pi/2) q[1],q[3];
h q[3];
cu1(-pi/4) q[1],q[4];
cu1(-pi/2) q[3],q[4];
h q[4];
cx q[4],q[1];
ccx q[4],q[1],q[2];
cx q[2],q[3];
ccx q[4],q[1],q[2];
cx q[4],q[1];
x q[2];
cx q[4],q[3];
h q[4];
x q[4];
measure q[3] -> c[3];
measure q[1] -> c[2];
measure q[2] -> c[1];
measure q[0] -> c[0];
```

**The circuit code for *n*=3**
```
#9 qubit and 217 one & two qubit
gates
#qubit 0 Anc
#qubit 1, 2 and 4, 6 corresponds to Reg. E and C
respectively
#qubit 3, 5, 7 Reg. B
#qubit 8 Anc.2
qreg q[9];
creg c[4];

#state b preparation
h q[5];
h q[7];
x q[8];
x q[5];
x q[7];
h q[8];
ccx q[7],q[5],q[3];
x q[5];
x q[7];
cry(pi/4) q[5],q[3];
cx q[7],q[5];
cry(-pi/4) q[5],q[3];
cx q[7],q[5];
cry(pi/4) q[7],q[3];
x q[5];
cry(pi/4) q[5],q[3];
cx q[7],q[5];
cry(-pi/4) q[5],q[3];
cx q[7],q[5];
cry(pi/4) q[7],q[3];
x q[5];
x q[7];
cry(pi/4) q[5],q[3];
cx q[7],q[5];
cry(-pi/4) q[5],q[3];
cx q[7],q[5];
cry(pi/4) q[7],q[3];
x q[7];
#state b preparation done

#sin transform
cx q[8],q[7];
cx q[8],q[5];
cx q[8],q[3];
ccx q[3],q[8],q[4];
ccx q[4],q[5],q[6];
cx q[6],q[7];
ccx q[4],q[5],q[6];
cx q[4],q[5];
ccx q[3],q[8],q[4];
cx q[8],q[3];
h q[8];
cp(pi/2) q[7],q[8];
cp(pi/4) q[5],q[8];
cp(pi/8) q[3],q[8];
h q[7];
cp(pi/2) q[5],q[7];
cp(pi/4) q[3],q[7];
h q[5];
cp(pi/2) q[3],q[5];
h q[3];
swap q[3],q[8];
swap q[5],q[7];
cx q[8],q[3];
ccx q[3],q[8],q[4];
cx q[4],q[5];
ccx q[4],q[5],q[6];
cx q[6],q[7];
ccx q[4],q[5],q[6];
ccx q[3],q[8],q[4];
```

```
cx q[8],q[3];
cx q[8],q[5];
barrier q[5];
cx q[8],q[7];
#sin transform done

#controlled R_y module
x q[5];
x q[7];
cry(pi/4) q[7],q[6];
cry(pi/4) q[7],q[4];
cx q[3],q[7];
cry(-pi/4) q[7],q[6];
cry(-pi/4) q[7],q[4];
cx q[3],q[7];
cry(pi/4) q[3],q[6];
x q[7];
cry(pi/4) q[3],q[4];
cry(pi/8) q[5],q[6];
x q[7];
cry(pi/8) q[5],q[4];
cx q[3],q[5];
cry(-pi/8) q[5],q[6];
cry(-pi/8) q[5],q[4];
cx q[3],q[5];
cry(pi/8) q[3],q[6];
cry(pi/8) q[3],q[4];
x q[5];
cry(pi/8) q[3],q[6];
cry(pi/8) q[3],q[4];
cx q[7],q[5];
cry(pi/4) q[5],q[2];
cry(pi/4) q[5],q[1];
cx q[3],q[5];
cry(-pi/4) q[5],q[2];
cry(-pi/4) q[5],q[1];
cx q[3],q[5];
cry(pi/4) q[3],q[2];
cx q[7],q[5];
cry(pi/4) q[3],q[1];
cry(pi/6) q[5],q[6];
x q[7];
cry(pi/4) q[3],q[2];
cry(pi/6) q[5],q[4];
x q[7];
cry(pi/4) q[3],q[1];
x q[3];
cx q[3],q[5];
cry(-pi/6) q[5],q[6];
cry(-pi/6) q[5],q[4];
cx q[3],q[5];
cry(pi/6) q[3],q[6];
cry(pi/6) q[3],q[4];
cry(pi/4) q[7],q[2];

cry(pi/4) q[7],q[1];
ccx q[3],q[5],q[7];
cry(-pi/4) q[7],q[2];
cry(-pi/4) q[7],q[1];
ccx q[3],q[5],q[7];
cry(pi/8) q[5],q[2];
x q[7];
cry(pi/8) q[5],q[1];
cry(pi/6) q[7],q[6];
cx q[3],q[5];
cry(-pi/8) q[5],q[2];
cry(-pi/8) q[5],q[1];
cx q[3],q[5];
cry(pi/8) q[3],q[2];
cry(pi/8) q[3],q[1];
cry(pi/8) q[5],q[2];
cry(pi/8) q[5],q[1];
cx q[3],q[5];
cry(-pi/8) q[5],q[2];
cry(-pi/8) q[5],q[1];
cx q[3],q[5];
cry(pi/8) q[3],q[2];
x q[5];
cry(pi/8) q[3],q[1];
cry(pi/6) q[7],q[4];
x q[3];
x q[3];
cry(pi/6) q[7],q[2];
cry(pi/6) q[7],q[1];
ccx q[3],q[5],q[7];
cry(-pi/6) q[7],q[6];
cry(-pi/6) q[7],q[4];
cry(-pi/6) q[7],q[2];
cry(-pi/6) q[7],q[1];
ccx q[3],q[5],q[7];
cry(pi/12) q[5],q[6];
cx q[8],q[7];
cry(pi/12) q[5],q[4];
cry(pi/12) q[5],q[2];
cry(pi/12) q[5],q[1];
cx q[3],q[5];
cry(-pi/12) q[5],q[6];
cry(-pi/12) q[5],q[4];
cry(-pi/12) q[5],q[2];
cry(-pi/12) q[5],q[1];
cx q[3],q[5];
cry(pi/12) q[3],q[6];
cry(pi/12) q[3],q[4];
x q[5];
cry(pi/12) q[3],q[2];
cry(pi/12) q[3],q[1];
crz(pi) q[1],q[0];
x q[3];
#controlled R_y module done
```

```
cry(pi/2) q[1],q[0];
crz(pi) q[2],q[0];
cry(pi/2) q[2],q[0];
ccx q[4],q[6],q[0];
cry(-pi/2) q[2],q[0];
crz(-pi/2) q[2],q[0];
ccx q[4],q[6],q[0];
crz(-pi/2) q[2],q[0];
cry(-pi/2) q[1],q[0];
crz(-pi/2) q[1],q[0];
crz(pi) q[2],q[0];
cry(pi/2) q[2],q[0];
ccx q[4],q[6],q[0];
cry(-pi/2) q[2],q[0];
crz(-pi/2) q[2],q[0];
ccx q[4],q[6],q[0];
crz(-pi/2) q[2],q[0];
x q[6]; #X module
crz(-pi/2) q[1],q[0];
x q[4]; #X module
cx q[8],q[5];
rx(pi) q[0];
cx q[8],q[3];
x q[0];
ccx q[3],q[8],q[4];
ccx q[4],q[5],q[6];
cx q[6],q[7];
ccx q[4],q[5],q[6];
cx q[4],q[5];
ccx q[3],q[8],q[4];
cx q[8],q[3];
swap q[5],q[7];
swap q[3],q[8];
h q[3];
cp(-pi/2) q[3],q[5];
h q[5];
cp(-pi/4) q[3],q[7];
cp(-pi/2) q[5],q[7];
h q[7];
cp(-pi/8) q[3],q[8];
cp(-pi/4) q[5],q[8];
cp(-pi/2) q[7],q[8];
h q[8];
cx q[8],q[3];
ccx q[3],q[8],q[4];
cx q[4],q[5];
ccx q[4],q[5],q[6];
cx q[6],q[7];
ccx q[4],q[5],q[6];
ccx q[3],q[8],q[4];
cx q[8],q[3];
cx q[8],q[5];
cx q[8],q[7];
h q[8];
x q[8];
measure q[7] -> c[3];
measure q[5] -> c[2];
measure q[3] -> c[1];
measure q[0] -> c[0];
```

**The Qiskit code for ideal circuit simulation without noise**

```
# preparation for execution
from qiskit import QuantumRegister, ClassicalRegister, QuantumCircuit, execute,Aer
from qiskit.visualization import plot_histogram
from numpy import pi

#Put Qiskit circuit code for n=2 or n=3

#execute the circuit
backend=Aer.get_backend('qasm_simulator')
job=execute(circuit,backend,shots=16384)
result = job.result()
counts = result.get_counts(circuit)
plot_histogram(counts).show()
```

**The $n$=2 Qiskit code for device backend noise model mimicked IBMQ_SANTIAGO**

```
#preparation for execution
from qiskit import QuantumRegister, ClassicalRegister, QuantumCircuit, execute
from qiskit.providers.aer import QasmSimulator
from qiskit.visualization import plot_histogram
from qiskit.test.mock import FakeSantiago
from numpy import pi

#Put Qiskit circuit code for n=2

#execute the circuit
device_backend = FakeSantiago()
Santiago_simulator=QasmSimulator.from_backend(device_backend)
result=execute(circuit,Santiago_simulator).result()
counts =result.get_counts(circuit)
plot_histogram(counts).show()
```

**The $n$=2 Qiskit code for executing circuit on real quantum computer IBMQ_SANTIAGO**

```
#preparation for execution
from qiskit import QuantumRegister, ClassicalRegister, QuantumCircuit, execute
```

```
from qiskit.visualization import plot_histogram
from numpy import pi
from qiskit import IBMQ
provider = IBMQ.load_account()
backend=provider.get_backend('ibmq_santiago')

#Put Qiskit circuit code for n=2

#execute the circuit
job=execute(circuit,backend,shots=8192)
result = job.result()
counts = result.get_counts(circuit)
plot_histogram(counts).show()
```

**The Qiskit code of adding four common kinds of noise**

```
#preparation for execution
from qiskit import QuantumRegister, ClassicalRegister, QuantumCircuit, execute
from qiskit.providers.aer import QasmSimulator
from qiskit.providers.aer.noise import NoiseModel
from qiskit.providers.aer.noise import QuantumError
from qiskit.providers.aer.noise import amplitude_damping_error, phase_damping_error, depolarizing_error, pauli_error
from qiskit.visualization import plot_histogram
from numpy import pi

#Put Qiskit circuit code for n=2 or n=3

#select the type of noise to add and set the noise intensity
p = 1e-04*pow(700,1)#set intensity noise
error=amplitude_damping_error(p)
error = phase_damping_error(p)
error = depolarizing_error(p, 1)
error = pauli_error([('X',p_error), ('I', 1 - p_error)])

#put selected noise on gates of the circuit
error_gatec = error.tensor(error)
error_gateccx=error.tensor(error).tensor(error)
noise_model = NoiseModel()
noise_model.add_all_qubit_quantum_error(error, ["h"])
noise_model.add_all_qubit_quantum_error(error, ["x"])
noise_model.add_all_qubit_quantum_error(error_gatec, ["cx"])
noise_model.add_all_qubit_quantum_error(error_gatec, ["cry"])
noise_model.add_all_qubit_quantum_error(error_gatec, ["cu1"])
noise_model.add_all_qubit_quantum_error(error_gateccx, ["ccx"])
noise_model.add_all_qubit_quantum_error(error, "measure")

#execute the noisy circuit
noise_simulator=QasmSimulator(noise_model=noise_model)
job = execute(circuit, noise_simulator)
noisy_result = job.result()
noisy_counts=noisy_result.get_counts(0)
plot_histogram(noisy_counts).show()
```

**Appendix B Date**

The detailed data on Figure 9 in the table below.

Table.1 The deviation for n=2 under four types of noise

| $i$ | AD | PD | DE | BF |
| --- | --- | --- | --- | --- |
| 1 | 0.0341 | 0.0206 | 0.0175 | 0.0161 |
| 2 | 0.0177 | 0.0275 | 0.0350 | 0.0293 |
| 3 | 0.0543 | 0.0396 | 0.0848 | 0.1001 |
| 4 | 0.0696 | 0.0310 | 0.0697 | 0.1184 |
| 5 | 0.1588 | 0.0196 | 0.1288 | 0.2381 |
| 6 | 0.2701 | 0.0412 | 0.2591 | 0.3782 |
| 7 | 0.4608 | 0.0606 | 0.4171 | 0.5390 |
| 8 | 0.6660 | 0.1504 | 0.5901 | 0.6528 |

| i | | | | |
|---|---|---|---|---|
| 9 | 0.8118 | 0.2412 | 0.7074 | 0.7047 |

Table.2 The deviation for n=3 under four types of noise

| $i$ | AD | PD | DE | BF |
|---|---|---|---|---|
| 1 | 0.0745 | 0.0625 | 0.0310 | 0.0778 |
| 2 | 0.1205 | 0.0498 | 0.1069 | 0.1580 |
| 3 | 0.1363 | 0.0372 | 0.1471 | 0.2116 |
| 4 | 0.2031 | 0.0842 | 0.2183 | 0.3565 |
| 5 | 0.3497 | 0.0900 | 0.3481 | 0.5365 |
| 6 | 0.4702 | 0.1781 | 0.4811 | 0.5875 |
| 7 | 0.5709 | 0.1993 | 0.6183 | 0.6690 |
| 8 | 0.6358 | 0.3257 | 0.5904 | 0.6736 |
| 9 | 0.7105 | 0.4825 | 0.6418 | 0.6212 |

**References**

1. See IBM Quantum. https://quantum-computing.ibm.com/ (2021) for detailed information
2. See https://qiskit.org/documentation/getting_started.html for detailed information